\documentclass[reprint,
 amsmath,amssymb,
prb,
]{revtex4-1}

\usepackage{graphicx}
\usepackage{dcolumn}
\usepackage{longtable}
\usepackage{lipsum}
\usepackage{color}
\usepackage[hypertex]{hyperref}
\usepackage{bm}
\usepackage{braket}
\begin{document}


\title{ Superconductivity in Cu-doped Bi$_2$Se$_3$ with potential disorder
}
\author{Takumi Sato$^{1}$}
\author{Yasuhiro Asano$^{1,2}$}
\affiliation{$^{1}$ Department of Applied Physics,
Hokkaido University, Sapporo 060-8628, Japan\\
$^{2}$Center of Topological Science \& Technology,
Hokkaido University, Sapporo 060-8628, Japan\\
}%

\date{\today}

\begin{abstract}
We study the effects of random nonmagnetic impurities on 
superconducting transition temperature $T_c$ 
in a Cu-doped Bi$_2$Se$_3$, 
for which four types of pair potentials have been proposed.
Although all the candidates belong to $s$-wave symmetry,
two orbital degree of freedom in electronic structures enriches
the symmetry variety of a Cooper pair such as even-orbital-parity and odd-orbital-parity.
We consider realistic electronic structures of Cu-doped Bi$_2$Se$_3$ by using tight-binding
Hamiltonian on a hexagonal lattice and consider effects of impurity scatterings through 
the self-energy of the Green's function within the Born approximation.
We find that even-orbital-parity spin-singlet superconductivity is basically robust even
in the presence of impurities. The degree of the robustness depends on the electronic
structures in the normal state and on the pairing symmetry in orbital space. 
On the other hand, two odd-orbital-parity spin-triplet order parameters are
always fragile in the presence of potential disorder.
\end{abstract}


\maketitle

\section{Introduction}

The robustness of superconductivity in the presence of nonmagnetic impurities 
depends on symmetry of the pair potential. 
The transition temperature $T_c$ is insensitive 
to the impurity concentration in a spin-singlet $s$-wave superconductor~\cite{AGD_book,Abrikosov_sovphys1959,Anderson_jpcs1959}.
In a cuprate superconductor, on the other hand, $T_c$ of a spin-singlet 
$d$-wave superconductivity is suppressed drastically by the impurity 
scatterings~\cite{Sun_prb1995}.
The pair potential of an unconventional superconductor 
changes its sign on the Fermi surface depending on the direction of a quasiparticle's momenta. 
The random impurity scatterings make the motion of a quasiparticle be isotropic 
in both real and momentum spaces. Such a diffused quasiparticle feels the 
pair potential averaged over the directions of momenta. 
The resulting pair potential is finite for an $s$-wave symmetry, whereas it 
is zero for unconventional pairing symmetries.
Thus, unconventional superconductivity is fragile under the potential disorder.

Previous papers~\cite{allen:review1982,Golubov_prb1997,Efremov_prb2011,Asano_prb2018,Asano_njphys2018} 
showed that $s$-wave superconductivity is not always robust against the nonmagnetic impurity 
scatterings in multiband (multiorbital) superconductors. 
The interorbital impurity scatterings decrease $T_c$, which is a common conclusion 
of all the theoretical studies.  
The two-band models considered in these papers, however, 
are too simple to discuss the effects of impurities on $T_c$ in real materials
such as iron pnictides~\cite{Kamihara_jacm2008,Kuroki_prl2008}, 
MgB$_2$~\cite{Nagamatsu_nature2001,Choi_nature2002}, 
and Cu-doped Bi$_2$Se$_3$~\cite{Hor_prl2010,Fu_prl2010}.
The robustness of multiband superconductivity under the potential disorder may depend 
on electronic structures near the Fermi level.
In iron pnictides and MgB$_2$, 
two electrons in the same conduction band form a 
Cooper pair~\cite{Kuroki_prl2008,Choi_nature2002}.
The impurity effect on such an intraband pair has been studied by taking 
realistic electronic structures into account~\cite{Onari_prl2009}.
In the case of Cu-doped Bi$_2$Se$_3$, four types of pair potentials $\Delta_1-\Delta_4$ 
have been proposed as a promising candidate of order parameter~\cite{Fu_prl2010}. 
Among them, an interorbital pairing order has attracted much attention 
as a topologically nontrivial superconductivity~\cite{Fu_prl2010,Sasaki_prl2011}.
Unfortunately, the possibility of such a topological superconductivity 
under the potential disorder has never been studied yet.
We address this issue.

In this paper, we study the effects of impurities on $T_c$ of Cu-doped Bi$_2$Se$_3$. 
We describe electronic structures near the Fermi level by taking into account 
two $p$ orbitals in Bi$_2$Se$_3$ and the hybridization 
between them~\cite{Zhang_natphys2009,Liu_prb2010}. 
According to the theoretical proposal~\cite{Fu_prl2010}, 
we consider four types of $s$-wave pair potential on such orbital based electronic structures. 
The effects of impurities on $T_c$ are estimated 
through the impurity self-energy within the Born approximation.
The transition temperature is calculated by solving the gap equation numerically
and is plotted as a function of impurity concentration $n_{\mathrm{imp}}$.
We will show that the relation between $T_c$ and $n_{\mathrm{imp}}$ 
depends sensitively on the types of pair potentials.
Superconductivity with an intraorbital pair potential $\Delta_1$ 
is robust even in the dirty regime.
This conclusion is consistent with that at a limiting case of 
previous studies~\cite{Golubov_prb1997,Efremov_prb2011,Asano_prb2018}.
There are two kinds of interorbital pairing order: 
even-orbital symmetry and odd-orbital symmetry.
We find that $T_c$ of an even-interorbital superconductivity $\Delta_3$
decreases slowly with the increase of $n_{\mathrm{imp}}$ and vanishes in the dirty limit.
The results for $\Delta_3$ disagree with those in a simple two-band model~\cite{Asano_njphys2018}
because the robustness of $\Delta_3$ depends sensitively on electronic structures.
Finally, the odd-interorbital pairing orders ($\Delta_2$ and $\Delta_4$) 
vanish at a critical value of 
the impurity concentration, which agrees well with the results of a idealistic two-band model~\cite{Asano_njphys2018}. 
Thus we conclude that odd-orbital pair potential is fragile irrespective of electronic structures.

This paper is organized as follows.
In Sec.~\ref{sectoion:model}, we describe the effective Hamiltonian near the Fermi level 
in Cu-doped Bi$_2$Se$_3$ and four types of pair potentials in its superconducting state.
The anomalous Green's function and the gap equation for each pair potential 
in the clean limit are obtained by solving the Gor'kov equation.
In Sec.~\ref{section:imp}, we introduce the random impurity potential 
and discuss the effects of impurities on $T_c$ within the Born approximation.
The conclusion is given in Sec.~\ref{section:conclusion}.
Throughout this paper, we use the units of $k_\mathrm{B}=\hbar=1$, where $k_{\mathrm{B}}$
is the Boltzmann constant.
The symbol $\bar{\cdots}$, $\check{\cdots}$, and $\hat{\cdots}$ represent
$8 \times 8$, $4 \times 4$, and $2 \times 2$ matrices, respectively.

\section{Clean limit}
\label{sectoion:model}
\subsection{Model}
\begin{figure}[tbh]
\begin{center}
\includegraphics[clip,scale=0.35]{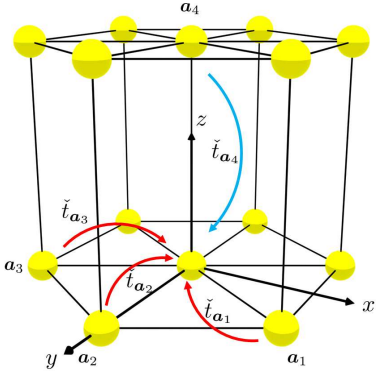}
\end{center}
\caption{
The simplified lattice structure of a Cu-doped Bi$_2$Se$_3$. 
The arrow indicates the hopping.
}
\label{fig:hexagonal lattice}
\end{figure}

For constructing an effective model of the normal state, 
we start with the tight-binding Hamiltonian on a 
hexagonal lattice as shown in 
Fig.~\ref{fig:hexagonal lattice}~\cite{Hashimoto_jpsj2013}. 
Strictly speaking, the crystal structure of 
Bi$_2$Se$_3$ is rhombohedral~\cite{Zhang_natphys2009,Liu_prb2010}.
The simplification does not affect the low energy physics.
We assume that an intercalated copper atom supplies electrons and makes 
a topological insulator Bi$_2$Se$_3$ be metallic~\cite{Wray_natphys2010}.
In the hexagonal lattice, the primitive lattice vectors are 
$\left( \sqrt{3}a/2 , a/2 , 0 \right)$, $\left( 0 , a , 0 \right)$, $\left( 0 , 0 , c \right)$ 
where $a$ and $c$ are the lattice constants in the $xy$ plane and along the $z$ axis, respectively.
We define the nearest neighbor vectors $\bm{a}_1=\left( \sqrt{3}a/2 , a/2 , 0 \right)$, $\bm{a}_2=\left( 0 , a , 0 \right)$, 
$\bm{a}_3=\left( -\sqrt{3}a/2 , a/2 , 0 \right)$, and $\bm{a}_4=\left( 0 , 0 , c \right)$.
The tight-binding Hamiltonian in real space can be written as~\cite{Hashimoto_jpsj2013,Mao_prb2011}
\begin{align}
H_N &= \sum_{\bm{R}} \bm{\psi}^{\dag}_{\bm{R}} \check{\varepsilon} \bm{\psi}_{\bm{R}}
+ \sum_{\bm{R},i} \bm{\psi}^{\dag}_{\bm{R}} \check{t}_{\bm{a}_i} \bm{\psi}_{\bm{R}+\bm{a}_i} + \mathrm{H.c.} , \\
\bm{\psi}_{\bm{R}} &= \left[ \psi_{P1_z^+,\uparrow}(\bm{R}) \, , \, \psi_{P2_z^-,\uparrow}(\bm{R}) \, , \,
                             \psi_{P1_z^+,\downarrow}(\bm{R}) \, , \, \psi_{P2_z^-,\downarrow}(\bm{R}) \right]^\mathrm{T} ,
\end{align}
where $\psi^{\dag}_{\sigma,s} \ \left(\psi_{\sigma,s}\right)$ is the creation (annihilation) operator
of an electron at the orbital $\sigma$ ($=P1_{z}^+$ or $P2_{z}^-$) with spin $s$ ($=\uparrow$ or $\downarrow$). We consider only the nearest neighbor hopping on the hexagonal lattice in the $xy$ 
plane and that along the $z$ axis.
An orbital $P1_{z}^+$ ($P2_{z}^-$) mainly consists of $p_z$ orbital of a Bi (Se) atom.
The matrix element of hopping $\check{t}_{\bm{a}_i}$ ($i=1-4$) is described as
\begin{align}
\bra{\bm{R},\sigma,s} H \ket{\bm{R}+\bm{a}_i,\sigma',s'} .
\end{align}
The nearest neighbor hopping elements are illustrated in Fig.~\ref{fig:hexagonal lattice}.
In momentum space, the tight-binding Hamiltonian is described as 
\begin{align}
\check{H}_N (\bm{k}) &= \check{\varepsilon} 
                      + \sum_{i} \check{t}_{\bm{a}_i} e^{i\bm{k} \cdot \bm{a}_i} + \mathrm{H.c.} .
\end{align}
The matrix structures of $\check{t}_{\bm{a}_i}$ are given in 
Appendix~\ref{section:tb hamiltonian}.
The tight-binding Hamiltonian can be written as
\begin{align}
\check{H}_N(\bm{k}) &= c_{\bm{k}} \hat{s}_0 \hat{\sigma}_0 + m_{\bm{k}} \hat{s}_0 \hat{\sigma}_3 \nonumber \\
\label{hn}
&\hspace{3.5pc} + V_z \hat{s}_0 \hat{\sigma}_2 + (V_y \hat{s}_1 - V_x \hat{s}_2) \hat{\sigma}_1 , \\
c_{\bm{k}} &= -\mu + c_1 \alpha_1 (\bm{k}) + c_2 \alpha_2 (\bm{k}) , \\
m_{\bm{k}} &= m_0 + m_1 \alpha_1 (\bm{k}) + m_2 \alpha_2 (\bm{k}) , \\
V_{x,y} &= v \alpha_{x,y} (\bm{k}) , \\
V_z &= v_z \alpha_z(\bm{k}) , 
\end{align}
where $\alpha_i(\bm{k})$ ($i=1,2,x,y,z$) is
\begin{align}
\label{a1}
\alpha_1(\bm{k}) &= \frac{2}{c^2} \left( 1-\cos k_z c \right) , \\
\label{a2}
\alpha_2(\bm{k}) &= \frac{4}{3a^2} \left( 3-2\cos \frac{\sqrt{3} k_x a}{2} \cos \frac{k_y a}{2} - \cos k_y a \right) , \\
\label{ax}
\alpha_x(\bm{k}) &= \frac{2}{\sqrt{3} a} \sin \frac{\sqrt{3}k_x a}{2} \cos \frac{k_y a}{2} , \\
\label{ay}
\alpha_y(\bm{k}) &= \frac{2}{3a} \left( \cos \frac{\sqrt{3} k_x a}{2} \sin \frac{k_y a}{2} + \sin k_y a \right) , \\
\label{az}
\alpha_z(\bm{k}) &= \frac{1}{c} \sin k_z c .
\end{align} 
We define the Pauli matrices $\hat{s}_j$ in spin space, $\hat{\sigma}_j$ in orbital space, 
and $\hat{\tau}_j$ in particle-hole space for $j=1-3$. 
The unit matrix in these spaces are $\hat{s}_0$, $\hat{\sigma}_0$, and $\hat{\tau}_0$.
In Eq.~\eqref{hn}, the hopping 
in the $z$ direction ($\check{t}_{\bm{a}_4}$) causes
the orbital hybridization term $V_z$ and 
the hopping in the $xy$ plane
($\check{t}_{\bm{a}_{1}}, \, \check{t}_{\bm{a}_{2}}, \, \check{t}_{\bm{a}_{3}}$)
causes the spin-orbit interaction term $V_{x,y}$.
When we expand the trigonometric functions around the $\Gamma$ point, 
the tight-binding Hamiltonian $\check{H}_N (\bm{k})$ 
corresponds to 
$\bm{k} \cdot \bm{p}$ Hamiltonian of Bi$_2$Se$_3$~\cite{Zhang_natphys2009,Liu_prb2010}.

The superconducting state in Cu$_x$Bi$_2$Se$_3$ is described by a Hamiltonian
\begin{align}
&\mathcal{H}^{(0)} = \sum_{\bm{k}} \bm{\Psi}^{\dag} (\bm{k}) \bar{H}^{(0)}_{\bm{k}} \bm{\Psi} (\bm{k}) , \quad
\bm{\Psi} (\bm{k}) = 
\left[ \begin{array}{c}
\bm{\psi}_e (\bm{k}) \\
\bm{\psi}_h (\bm{k})
\end{array} \right] , \\
&\bm{\psi}_e (\bm{k}) = 
\left[ \begin{array}{c}
\psi_{P1_z^+,\uparrow} (\bm{k}) \\
\psi_{P2_z^-,\uparrow} (\bm{k}) \\
\psi_{P1_z^+,\downarrow} (\bm{k}) \\
\psi_{P2_z^-,\downarrow} (\bm{k})
\end{array} \right], \;
\bm{\psi}_h (\bm{k}) =
\left[ \begin{array}{c}
\psi^{\dag}_{P1_z^+,\uparrow} (-\bm{k}) \\
\psi^{\dag}_{P2_z^-,\uparrow} (-\bm{k}) \\
\psi^{\dag}_{P1_z^+,\downarrow} (-\bm{k}) \\
\psi^{\dag}_{P2_z^-,\downarrow} (-\bm{k})
\end{array} \right] , \\
\label{hbdg}
&\bar{H}^{(0)}_{\bm{k}} = \left( 
\begin{array}{cc}
\check{H}_N(\bm{k}) & \check{\Delta}_{\lambda} \\
\check{\Delta}^{\dag}_{\lambda} & -\check{H}^{\ast}_N(-\bm{k})
\end{array}
\right) .
\end{align}
According to the previous proposal~\cite{Fu_prl2010}, we consider 
four types of momentum-independent pair potential
defined by
\begin{align}
\label{delta1}
\Delta_1 &= \frac{g_1}{N} \sum_{\bm{k}}
\langle \psi_{P1_{z}^+,\uparrow} (\bm{k}) \psi_{P1_{z}^+,\downarrow} (-\bm{k}) \rangle \nonumber\\
&= \frac{g_1}{N} \sum_{\bm{k}}
\langle \psi_{P2_{z}^-,\uparrow} (\bm{k}) \psi_{P2_{z}^-,\downarrow} (-\bm{k}) \rangle,\\
\label{delta2}
\Delta_2 &= \frac{g_2}{N} \sum_{\bm{k}}
\langle \psi_{P1_{z}^+,\uparrow} (\bm{k}) \psi_{P2_{z}^-,\downarrow} (-\bm{k}) \rangle \nonumber\\
&=-\frac{g_2}{N} \sum_{\bm{k}}
\langle \psi_{P2_{z}^-,\uparrow} (\bm{k}) \psi_{P1_{z}^+,\downarrow} (-\bm{k}) \rangle,\\
\label{delta3}
\Delta_3 &= \frac{g_3}{N} \sum_{\bm{k}}
\langle \psi_{P1_{z}^+,\uparrow} (\bm{k}) \psi_{P2_{z}^-,\downarrow} (-\bm{k}) \rangle \nonumber\\
&= \frac{g_3}{N} \sum_{\bm{k}}
\langle \psi_{P2_{z}^-,\uparrow} (\bm{k}) \psi_{P1_{z}^+,\downarrow} (-\bm{k}) \rangle,\\
\label{delta4}
\Delta_4 &= \frac{g_4}{N} \sum_{\bm{k}}
\langle \psi_{P1_{z}^+,\uparrow} (\bm{k}) \psi_{P2_{z}^-,\uparrow} (-\bm{k}) \rangle \nonumber\\
&=-\frac{g_4}{N} \sum_{\bm{k}}
\langle \psi_{P2_{z}^-,\uparrow} (\bm{k}) \psi_{P1_{z}^+,\uparrow} (-\bm{k}) \rangle,
\end{align}
where $g_{\lambda}>0 \ (\lambda=1-4)$ represents the attractive interaction between two electrons.
Generally speaking, the pair correlation function can be represented as
\begin{align}
f_{s,\sigma; s^\prime, \sigma^\prime}(\boldsymbol{k}) 
=\left\langle  \psi_{s,\sigma}(\boldsymbol{k})\,  \psi_{s^\prime,\sigma^\prime}(-\boldsymbol{k})
\right\rangle,
\end{align}
where we assume a spatially uniform equal-time Cooper pair. 
The momentum-symmetry is even-parity $s$-wave symmetry, which is 
a common property among the four candidates in a Cu-doped Bi$_2$Se$_3$.
Because of the Fermi-Dirac statistics of electrons, the pairing correlation 
obeys
\begin{align}
f_{s,\sigma; s^\prime, \sigma^\prime}(\boldsymbol{k}) 
= - f_{s^\prime, \sigma^\prime; s, \sigma}(\boldsymbol{k}). 
\end{align}
The remaining symmetry options of the pairing function are orbitals and spins of 
a Cooper pair. Therefore, the pairing function must be either antisymmetric 
under $s \leftrightarrow s^\prime$ or antisymmetric 
under $\sigma \leftrightarrow \sigma^\prime$.

Both Eqs.~\eqref{delta1} and \eqref{delta3} belong to spin-singlet symmetry.
Thus the pairing functions belong to even-orbital parity.
In Eq.~\eqref{delta3}, a Cooper pair consists of two electrons in the different 
orbitals (interorbital pair):
one electron is in $P1_{z}^+$ orbital 
and the other is in $P2_{z}^-$ orbital. 
In Eq.~\eqref{delta1}, on the other hand, 
a Cooper pair consists of two electrons in the same orbital (intraorbital pair).
The pair potential in the $P1_{z}^+$ orbital and that in the $P2_{z}^-$ orbital 
have the same amplitude and the same sign. 

Both Eqs.~\eqref{delta2} and \eqref{delta4} represent the spin-triplet interorbital pairing correlations.
In these cases, the pair correlation belongs to odd-orbital-parity symmetry.
In addition to the symmetry options for Cooper pairing,
the pair potentials are classified by the irreducible representation of $D_{3d}$ point group. 
$\Delta_2$ and $\Delta_4$ can be distinguished from each other by the 
irreducible representation. 
The matrix form of pair potentials, the irreducible representation, spin symmetry, 
and orbital-parity of the pair potentials 
are summarized in Table~\ref{pair potentials}. 
Although Fu and Berg~\cite{Fu_prl2010} proposed a pair potential of
$\Delta (i\hat{s}_2) \hat{\sigma}_3$, it is unitary equivalent to Eq.~(\ref{delta1}) 
as long as the Hamiltonian $\bar{H}^{(0)}_{\bm{k}}$ preserves time-reversal symmetry~\cite{Asano_prb2018}. 
(See Appendix~\ref{section:eq between 1a & 1b} for details.)
They also considered a pair potential of
$\Delta \hat{s}_0 (i\hat{\sigma}_2)$ independently of Eq.~(\ref{delta4}). 
However, the behavior of $T_c$ under the potential disorder 
in the two pair potentials are the same with each other.
Thus, in this paper, we discuss effects of random impurity scatterings on 
superconducting states described by Eqs.~(\ref{delta1})-(\ref{delta4}).
We note that 
the orbital parity and the momentum parity are independent symmetry options of each other.
The former represents symmetry of correlation function under the commutation of two orbitals.
The latter is derived from inversion symmetry of the lattice structure.

\begin{center}
\begin{table}[tbh]
\caption{Symmetry classification of pair potentials. Equal-time pairing order 
parameter belongs to even-frequency symmetry.
A spin-singlet component is described by $i\hat{s}_2$. 
An opposite-spin-triplet and an equal-spin-triplet components are 
indicated by $\hat{s}_1$ and $\hat{s}_3$, respectively. 
}
\centering
\begin{tabular}{cccccc} \hline
Matrix & 
\begin{tabular}{c}
\end{tabular}
 Rep. & Frequency & Spin &
\begin{tabular}{c}
Momentum \\ parity
\end{tabular} &
\begin{tabular}{c}
Orbital \\ parity
\end{tabular} \\ \hline \hline
$\Delta_1 (i\hat{s}_2)\hat{\sigma}_0$  & $A_{1g}$ & Even & Singlet & Even 
& \begin{tabular}{c}
Even \\ (Intra)
\end{tabular} \\ \hline
$\Delta_2 \hat{s}_1 (i\hat{\sigma}_2)$ & $A_{1u}$ & Even & Triplet & Even 
& \begin{tabular}{c}
Odd \\ (Inter)
\end{tabular} \\ \hline
$\Delta_3 (i\hat{s}_2) \hat{\sigma}_1$ & $A_{2u}$ & Even & Singlet & Even
&\begin{tabular}{c}
Even \\ (Inter)
\end{tabular} \\ \hline
$\Delta_4 \hat{s}_3 (i\hat{\sigma}_2)$  & $E_{u}$ & Even & Triplet & Even
& \begin{tabular}{c}
Odd \\ (Inter)
\end{tabular} \\ \hline
\end{tabular}
\label{pair potentials}
\end{table}
\end{center}

\subsection{Gor'kov equation}
\label{subsec:gorkov eq}
The Matsubara Green's function is obtained by solving the Gor'kov equation,
\begin{gather}
\label{gorkov clean}
\left[ i \omega_n -  \check{H}^{(0)}(\bm{k}) \right] \bar{G}^{(0)}(\bm{k},i\omega_n) = 1 , \\
\bar{G}^{(0)}(\bm{k},i\omega_n) =
\left(
\begin{array}{cc}
\check{\mathcal{G}}^{(0)}(\bm{k},i\omega_n) & \check{\mathcal{F}}^{(0)}_{\lambda}(\bm{k},i\omega_n) \\
-{ \check{\mathcal{F}}^{(0) \ast}_{\lambda} } (-\bm{k},i\omega_n) & 
-{\check{\mathcal{G}}^{(0) \ast}}(-\bm{k},i\omega_n) 
\end{array}
\right) ,
\end{gather}
where $\omega_n=(2n+1)\pi T$ is a fermionic Matsubara frequency and $T$ is a temperature. 
To discuss the transition temperature,
we need to find the solutions of Eq.~\eqref{gorkov clean} 
within the first order of $\Delta$.
The results of the normal part 
\begin{widetext}
\begin{align}
\label{g0}
\check{\mathcal{G}}^{(0)}(\bm{k},i\omega_n) &= \frac{1}{X}
\left[ \left( i\omega_n - c_{\bm{k}} \right)\, \hat{s}_0 \hat{\sigma}_0 
              + m_{\bm{k}}\, \hat{s}_0 \hat{\sigma}_3
              + V_z\, \hat{s}_0 \hat{\sigma}_2 
              + (V_y \hat{s}_1 - V_x \hat{s}_2)\, \hat{\sigma}_1 \right] , \\
X(\bm{k},i\omega_n) &= (i\omega_n - c_{\bm{k}})^2 - m_{\bm{k}}^2 - V_x^2 - V_y^2-V_z^2 ,
\end{align}
are common for all the pair potentials because the normal Green's function 
does not include the pair potential at the lowest order.
The results of anomalous Green's function are given by,
\begin{align}			  
\check{\mathcal{F}}^{(0)}_1 (\bm{k},i\omega_n) &= \frac{\Delta_1}{Z}
\left[ -i\left(\omega^2_n+c^2_{\bm{k}} + m^2_{\bm{k}} 
+ V^2_x + V_y^2+V_z^2 \right)\, \hat{s}_2\hat{\sigma}_0 \right. \nonumber \\
\label{f1}
&\hspace{2pc} \left. + 2ic_{\bm{k}} m_{\bm{k}}\, \hat{s}_2 \hat{\sigma}_3 
                     + 2ic_{\bm{k}} V_z\, \hat{s}_2\hat{\sigma}_2
                     - 2c_{\bm{k}} V_y\, \hat{s}_3 \hat{\sigma}_1 
                     - 2ic_{\bm{k}}V_x\, \hat{s}_0\hat{\sigma}_1 \right] , \\
\check{\mathcal{F}}^{(0)}_2 (\bm{k},i\omega_n) &= \frac{\Delta_2}{Z}
\left[ -i\left(\omega^2_n+c^2_{\bm{k}}-m^2_{\bm{k}}+V_x^2+V_y^2
       +V_z^2 \right)\, \hat{s}_1\hat{\sigma}_2 \right. \nonumber \\
&\hspace{2pc} + 2m_{\bm{k}} V_y\, \hat{s}_0\hat{\sigma}_0 
              - 2c_{\bm{k}} V_y\, \hat{s}_0\hat{\sigma}_3 
              + 2im_{\bm{k}} V_x\, \hat{s}_3\hat{\sigma}_0 
              - 2ic_{\bm{k}} V_x\, \hat{s}_3\hat{\sigma}_3 \nonumber \\
\label{f2}
&\hspace{2pc} \left. + 2ic_{\bm{k}} V_z\, \hat{s}_1\hat{\sigma}_0 
                     - 2im_{\bm{k}} V_z\, \hat{s}_1\hat{\sigma}_3
                     + 2i\omega_n m_{\bm{k}}\, \hat{s}_1\hat{\sigma}_1 \right] , \\
\check{\mathcal{F}}^{(0)}_3 (\bm{k},i\omega_n) &= \frac{\Delta_3}{Z}
\left[ -i\left(\omega^2_n + c^2_{\bm{k}} - m^2_{\bm{k}}
 + V_x^2+V_y^2-V_z^2) \right)\, \hat{s}_2\hat{\sigma}_1 \right. \nonumber \\
&\hspace{2pc} + 2iV_x V_z\, \hat{s}_0\hat{\sigma}_2 
              + 2V_y V_z\, \hat{s}_3\hat{\sigma}_2
              - 2ic_{\bm{k}} V_x\, \hat{s}_0\hat{\sigma}_0 
              + 2im_{\bm{k}} V_x\, \hat{s}_0\hat{\sigma}_3  \nonumber \\
\label{f3}
&\hspace{2pc} \left. - 2c_{\bm{k}} V_y\, \hat{s}_3\hat{\sigma}_0 
                     + 2m_{\bm{k}} V_y\, \hat{s}_3\hat{\sigma}_3 
                     - 2i\omega_n m_{\bm{k}}\, \hat{s}_2\hat{\sigma}_2 
                     + 2i\omega_n V_z\, \hat{s}_2\hat{\sigma}_3 \right] , \\
\check{\mathcal{F}}^{(0)}_4 (\bm{k},i\omega_n) &= \frac{\Delta_4}{Z}
\left[ -i\left(\omega^2_n+c^2_{\bm{k}}-m^2_{\bm{k}}
       +V_x^2-V_y^2+V_z^2 \right)\, \hat{s}_3\hat{\sigma}_2 \right. \nonumber \\
&\hspace{2pc} - 2V_x V_y\, \hat{s}_0\hat{\sigma}_2 
              - 2im_{\bm{k}}\, V_x \hat{s}_1\hat{\sigma}_0 
              + 2ic_{\bm{k}} V_x\, \hat{s}_1\hat{\sigma}_3 
              + 2ic_{\bm{k}} V_z\, \hat{s}_3\hat{\sigma}_0 \nonumber \\
\label{f4}
&\hspace{2pc} \left. - 2im_{\bm{k}} V_z\, \hat{s}_3\hat{\sigma}_3 
                     - 2V_yV_z\, \hat{s}_2\hat{\sigma}_1
                     + 2i\omega_n m_{\bm{k}}\, \hat{s}_3\hat{\sigma}_1 
                     - 2\omega_n V_y\, \hat{s}_2\hat{\sigma}_3 \right] , 
\end{align}
with 
$Z(\bm{k},i\omega_n) = |X(\bm{k},i\omega_n)|^2$.
The $\hat{s}_2 \hat{\sigma}_0$ component in Eq.~\eqref{f1}, 
the $\hat{s}_1 \hat{\sigma}_2$ component in Eq.~\eqref{f2}, 
the $\hat{s}_2 \hat{\sigma}_1$ component in Eq.~\eqref{f3}, 
and the $\hat{s}_3 \hat{\sigma}_2$ component in Eq.~\eqref{f4} 
are linked to the pair potentials $\Delta_1$, $\Delta_2$,  $\Delta_3$, and $\Delta_4$, respectively. 
Therefore, the gap equations in the linear regime result in
\begin{align}
\Delta_1 &= -g_1 T \sum_{\omega_n}\frac{1}{N} \sum_{\bm{k}} \mathrm{Tr}
\left[ \check{\mathcal{F}}^{(0)}_1 (\bm{k},i\omega_n) \frac{(-i\hat{s}_2)\hat{\sigma}_0}{4} \right] \nonumber \\
\label{gapeq1}
&=g_1 T \sum_{\omega_n}\frac{1}{N} \sum_{\bm{k}}
\frac{\Delta_1}{Z(\bm{k},i\omega_n)} \left[ \omega_n^2 + c_{\bm{k}}^2 + m_{\bm{k}}^2
                                           + V_x^2 + V_y^2 + V_z^2 \right] , 
\end{align}
\begin{align}										  
\Delta_2 &= -g_2 T\sum_{\omega_n} \frac{1}{N} \sum_{\bm{k}} \mathrm{Tr} \left[ 
\check{\mathcal{F}}^{(0)}_2(\bm{k},i\omega_n) \frac{\hat{s}_1(-i\hat{\sigma}_2)}{4} \right] \nonumber \\
\label{gapeq2}
&= g_2 T\sum_{\omega_n} \frac{1}{N} \sum_{\bm{k}} \frac{\Delta_2}{Z(\bm{k},i\omega_n)} 
\left[ \omega_n^2 + c^2_{\bm{k}} -m_{\bm{k}}^2 
                       + V_x^2 + V_y^2 + V_z^2 \right] , 
\end{align}
\begin{align}
\Delta_3 &= -g_3 T \sum_{\omega_n}\frac{1}{N} \sum_{\bm{k}} \mathrm{Tr}
\left[ \check{\mathcal{F}}^{(0)}_3 (\bm{k},i\omega_n) \frac{(-i\hat{s}_2)\hat{\sigma}_1}{4} \right] \nonumber \\
\label{gapeq3}
&= g_3 T \sum_{\omega_n}\frac{1}{N} \sum_{\bm{k}}
\frac{\Delta_3}{Z(\bm{k},i\omega_n)} \left[ \omega_n^2 + c_{\bm{k}}^2 - m_{\bm{k}}^2
                       + V_x^2 + V_y^2 - V_z^2 \right] , 
\end{align}
\begin{align}
\Delta_4 &= -g_4 T\sum_{\omega_n} \frac{1}{N} \sum_{\bm{k}} \mathrm{Tr} \left[ 
\check{\mathcal{F}}^{(0)}_4(\bm{k},i\omega_n) \frac{\hat{s}_3(-i\hat{\sigma}_2)}{4} \right] \nonumber \\
\label{gapeq4}
&= g_4 T\sum_{\omega_n} \frac{1}{N} \sum_{\bm{k}} \frac{\Delta_4}{Z(\bm{k},i\omega_n)} 
\left[ \omega_n^2 + c_{\bm{k}}^2 - m_{\bm{k}}^2
                       + V_x^2 - V_y^2 + V_z^2 \right] .
\end{align}
\end{widetext}

Eqs.~\eqref{f1}, \eqref{f2}, \eqref{f3}, and \eqref{f4} 
show that the orbital hybridization ($V_z$), the spin-orbit interaction ($V_{x,y}$), 
and the asymmetry between the two orbitals ($m_{\bm{k}}$) generate 
various paring correlations which belong to different 
symmetry classes from that of the pair potential~\cite{BSchaffer:prb2013,Asano_prb2015}. 
Especially, we discuss briefly a role of odd-frequency pairing correlation in the gap equation.
For instance, the pairing correlation $\check{\mathcal{F}}^{(0)}_2$ includes 
$2i\omega_n m_{\bm{k}}\, \hat{s}_1\hat{\sigma}_1$ which describes a
spin-triplet even-orbital-parity component. Such an component must be odd-frequency 
symmetry because the pairing correlation function must be antisymmetric under the permutation of 
two electrons. 
In the gap equation, 
the odd-frequency pairing component decreases the numerator as shown in $-m_{\boldsymbol{k}}^2$ 
in Eq.~\eqref{gapeq2}. It has been pointed out that an odd-frequency pair
decreases the transition
temperature~\cite{Asano_prb2015}.
If we would be able to tune the parameters to delete more the odd-frequency components, 
the gap equation results in higher $T_c$.

\section{Effects of disorder}
\label{section:imp}
\begin{figure}[tbh]
\begin{center}
\includegraphics[clip,scale=0.35]{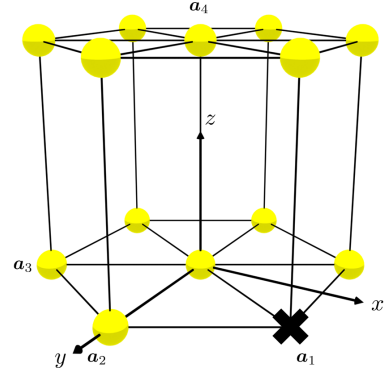}
\end{center}
\caption{
A model of the random potential in a Cu$_x$Bi$_2$Se$_3$. 
The cross mark denotes an impurity.
}
\label{fig:impurity}
\end{figure}

We consider the random nonmagnetic impurities described by
\begin{align}
\label{himp}
\bar{H}_{\mathrm{imp}}(\bm{R}) &=
V_{\mathrm{imp}}(\bm{R}) \  \hat{\tau}_3 \hat{s}_0 (\hat{\sigma}_0 + \hat{\sigma}_1) .
\end{align}
The schematic picture of potential disorder in a Cu$_x$Bi$_2$Se$_3$ is shown in Fig.~\ref{fig:impurity}.
We assume the impurity potential satisfies the following properties,
\begin{gather}
\label{vimp1}
\overline{V_{\mathrm{imp}}(\bm{R})} = 0 , \\
\label{vimp2}
\overline{V_{\mathrm{imp}}(\bm{R}) V_{\mathrm{imp}}(\bm{R}')} = n_{\mathrm{imp}} v^2_{\mathrm{imp}} \delta_{\bm{R},\bm{R}'} ,
\end{gather}
where $\overline{\cdots}$ means the ensemble average, $n_{\mathrm{imp}}$ is the density of the impurities, 
and $v_{\mathrm{imp}}$ is the strength of the impurity potential.
We also assume that the attractive interactions between two electrons are 
insensitive to the impurity potentials~\cite{Anderson_jpcs1959}.
We calculate the Green's function in the presence of the impurity potentials within the Born approximation. The Green's function is expanded up to the second order of the impurity potential. 
\begin{align}
\label{born1}
\bar{G}&(\bm{R}-\bm{R}',\omega_n) 
\approx \bar{G}^{(0)} (\bm{R}-\bm{R}',\omega_n) \nonumber\\
&+ \sum_{\bm{R}_1} \bar{G}^{(0)} (\bm{R}-\bm{R}_1,\omega_n) 
         \overline{\bar{H}_{\mathrm{imp}}(\bm{R}_1)} \bar{G}^{(0)} (\bm{R}_1-\bm{R}',\omega_n) 
		 \nonumber \\
&+ \sum_{\bm{R}_1,\bm{R}_2} \bar{G}^{(0)} (\bm{R}-\bm{R}_1,\omega_n)  \nonumber\\
 &\times  \overline{\bar{H}_{\mathrm{imp}}(\bm{R}_1) \bar{G}^{(0)} (\bm{R}_1-\bm{R}_2,\omega_n)    \bar{H}_{\mathrm{imp}}(\bm{R}_2)} \nonumber\\
&\times
\bar{G} (\bm{R}_2-\bm{R}',\omega_n) , \\
\label{born2}
\approx& \bar{G}^{(0)} (\bm{R}-\bm{R}',\omega_n) \nonumber \\
&+ n_{\mathrm{imp}} v^2_{\mathrm{imp}} \sum_{\bm{R}_1}
                   \bar{G}^{(0)} (\bm{R}-\bm{R}_1,\omega_n) \, \hat{\tau}_3\,  \hat{s}_0\,  \hat{\sigma}_0 
				   \nonumber\\
            &\times       \bar{G}^{(0)}(0,\omega_n)\,  \hat{\tau}_3 \, \hat{s}_0 \, \hat{\sigma}_0 
                   \bar{G}(\bm{R}_1 - \bm{R}',\omega_n) \nonumber \\                  
                 &+ n_{\mathrm{imp}} v^2_{\mathrm{imp}} \sum_{\bm{R}_1}
                   \bar{G}^{(0)} (\bm{R}-\bm{R}_1,\omega_n) \, \hat{\tau}_3\,  \hat{s}_0 \, \hat{\sigma}_1 
				   \nonumber\\ 
            &\times       \bar{G}^{(0)}(0,\omega_n) \, \hat{\tau}_3 \, \hat{s}_0 \, \hat{\sigma}_1 
                   \bar{G}(\bm{R}_1 - \bm{R}',\omega_n) .                            
\end{align}
We transform the Eq.~\eqref{born1} to \eqref{born2} 
by using the properties in Eqs.~\eqref{vimp1} and \eqref{vimp2}. 
In momentum space, Eq.~\eqref{born2} becomes
\begin{gather}
\bar{G}(\bm{k},i\omega_n) = \bar{G}^{(0)}(\bm{k},i\omega_n) + \bar{G}^{(0)}(\bm{k},i\omega_n)
                     \bar{\Sigma}_{\mathrm{imp}} \bar{G}(\bm{k},i\omega_n) , \\
\bar{\Sigma}_{\mathrm{imp}} = \bar{\Sigma}_{\mathrm{intra}} + \bar{\Sigma}_{\mathrm{inter}} , \\                    
\bar{\Sigma}_{\mathrm{intra}} = n_{\mathrm{imp}} v^2_{{\mathrm{imp}}} \hat{\tau}_3 \hat{s}_0 \hat{\sigma}_0
                            \frac{1}{N} \sum_{\bm{k}} 
                            \bar{G}^{(0)} (\bm{k},i\omega_n) \hat{\tau}_3 \hat{s}_0 \hat{\sigma}_0 , \\
\bar{\Sigma}_{\mathrm{inter}} = n_{\mathrm{imp}} v^2_{{\mathrm{imp}}} \hat{\tau}_3 \hat{s}_0 \hat{\sigma}_1
                            \frac{1}{N} \sum_{\bm{k}} 
                            \bar{G}^{(0)} (\bm{k},i\omega_n) \hat{\tau}_3 \hat{s}_0 \hat{\sigma}_1 ,                                
\end{gather}
where $\bar{\Sigma}_{\mathrm{intra}}$ and $\bar{\Sigma}_{\mathrm{inter}}$ are the self-energy 
due to the intraorbital impurity scatterings and that of the interorbital impurity scatterings, respectively. 
We describe the total self-energy as follows. 
\begin{gather}
\bar{\Sigma}_{\mathrm{imp}} = \bar{\Sigma}_{\mathrm{intra}} + \bar{\Sigma}_{\mathrm{inter}} 
=\left[
\begin{array}{cc}
\check{\Sigma}_g & \check{\Sigma}_{f_{\lambda}} \\
-\check{\Sigma}_{f_{\lambda}}^{\ast} & -\check{\Sigma}_g^{\ast}
\end{array}
\right] , \\
\label{sigmag}
\check{\Sigma}_g = n_{\mathrm{imp}} v^2_{\mathrm{imp}} [\check{g}^{(0)} 
                   + \hat{s}_0\hat{\sigma}_1 \check{g}^{(0)} \hat{s}_0\hat{\sigma}_1] , \\
\label{sigmaf}
\check{\Sigma}_{f_{\lambda}} =-n_{\mathrm{imp}} v^2_{\mathrm{imp}} [\check{f}_{\lambda}^{(0)} 
                   + \hat{s}_0\hat{\sigma}_1 \check{f}_{\lambda}^{(0)} \hat{s}_0\hat{\sigma}_1] ,
\end{gather}
where we denote the momentum summation of the Green's function
as $1/N \sum_{\bm{k}} \check{\mathcal{G}}^{(0)} (\bm{k},i\omega_n) = \check{g}^{(0)}$ 
and $1/N \sum_{\bm{k}} \check{\mathcal{F}}^{(0)} (\bm{k},i\omega_n) = \check{f}^{(0)}$. 
Therefore, the Gor'kov equation in the presence of the impurity potential is described by
\begin{gather}
\label{dirty gorkov eq}
\left[ i\omega_n - \bar{H}_0 (\bm{k}) - \bar{\Sigma}_{\mathrm{imp}} \right] \bar{G}(\bm{k},i\omega_n) = 1 , \\
\bar{G}(\bm{k},i\omega_n) =
\left(
\begin{array}{cc}
\check{\mathcal{G}}(\bm{k},i\omega_n) & \check{\mathcal{F}}_{\lambda}(\bm{k},i\omega_n) \\
-\check{\mathcal{F}}_{\lambda}^{\ast}(-\bm{k},i\omega_n) & -\check{\mathcal{G}}^{\ast} (-\bm{k},i\omega_n) 
\end{array}
\right) .
\end{gather}
The normal part of self-energy ($\Check{\Sigma}_g$) is calculated as follows.
\begin{align}
\label{normal part of se}
\check{\Sigma}_g 
=& \left[ -i\omega_n \eta_n + I_n \right]\, \hat{s}_0 \hat{\sigma}_0 , \\
\eta_n
=& n_{\mathrm{imp}} v^2_{\mathrm{imp}} \frac{1}{N} \sum_{\bm{k}} \frac{2}{Z} \hspace{10pc} \nonumber \\       
\hspace{3pc} &\times \left[\omega_n^2+c_{\bm{k}}^2+m_{\bm{k}}^2 + V_x^2+V_y^2+V_z^2 \right] , \\
I_n
=& n_{\mathrm{imp}} v^2_{\mathrm{imp}} \frac{1}{N} \sum_{\bm{k}}\frac{-2c_{\bm{k}}}{Z} \hspace{10pc} \nonumber \\
\hspace{3pc} &\times \left[ \omega_n^2+c_{\bm{k}}^2-m_{\bm{k}}^2 - V_x^2-V_y^2-V_z^2 \right] .
\end{align}
Within the first order of $\Delta$, the normal Green's function becomes
\begin{align}
\check{\mathcal{G}}(\bm{k},i\omega_n) &= \check{\mathcal{G}}^{(0)} (\bm{k},i\tilde{\omega}_n) 
                                |_{\mu \rightarrow \tilde{\mu}} , \\
\tilde{\omega}_n &= \omega_n (1+\eta_n) , \\
\label{mu shift}
\tilde{\mu} &= \mu - I_n .
\end{align}
The imaginary (real) part of the self-energy renormalizes the Matsubara frequency (chemical potential). 
The anomalous Green's function after summing up the momenta
is described as  
\begin{align}
\check{f}^{(0)}_1 &= \frac{1}{N} \sum_{\bm{k}} \frac{\Delta_1}{Z} \hspace{14pc} \nonumber \\
&\times \left[ -i\left( \omega_n^2+c_{\bm{k}}^2+m_{\bm{k}}^2
                        + V_x^2+V_y^2+V_z^2 \right) \, \hat{s}_2 \hat{\sigma}_0 \right. \nonumber \\
&\hspace{1pc} \left. + 2i cm \, \hat{s}_2 \hat{\sigma}_3 \right] , \\
\check{f}^{(0)}_2 &= \frac{1}{N} \sum_{\bm{k}} \frac{\Delta_2}{Z} \hspace{14pc} \nonumber \\
&\times \left[ -i\left( \omega_n^2+c_{\bm{k}}^2-m_{\bm{k}}^2
               + V_x^2+V_y^2+V_z^2 \right) \, \hat{s}_1 \hat{\sigma}_2 \right. \nonumber \\
\label{f_2}
&\hspace{1pc} \left. + 2i \omega_n m_{\bm{k}} \, \hat{s}_1 \hat{\sigma}_1 \right] , \\
\check{f}^{(0)}_3 &= \frac{1}{N} \sum_{\bm{k}} \frac{\Delta_3}{Z} \hspace{14pc} \nonumber \\
&\times \left[ -i\left( \omega_n^2+c_{\bm{k}}^2-m_{\bm{k}}^2
                        +V_x^2+V_y^2-V_z^2 \right)\, \hat{s}_2 \hat{\sigma}_1 \right. \nonumber \\
&\hspace{1pc} \left. - 2i \omega_n m \, \hat{s}_2 \hat{\sigma}_2 \right] , \\
\check{f}^{(0)}_4 &= \frac{1}{N} \sum_{\bm{k}} \frac{\Delta_4}{Z} \hspace{14pc} \nonumber \\
&\times \left[ -i\left( \omega_n^2+c_{\bm{k}}^2-m_{\bm{k}}^2
               + V_x^2-V_y^2+V_z^2 \right)\, \hat{s}_3 \hat{\sigma}_2 \right. \nonumber \\
\label{f_4}
&\hspace{1pc} \left. + 2i \omega_n m_{\bm{k}} \, \hat{s}_3 \hat{\sigma}_1 \right] .
\end{align}
By substituting these expressions into Eq.~\eqref{sigmaf}, we obtain the anomalous part of the self-energy 
for each pair potential.
\begin{align}
\label{sigmaf1}
\check{\Sigma}_{f_1}
&= \Delta_1  (i\hat{s}_2) \hat{\sigma}_0 \cdot \eta_n , \\
\label{sigmaf2}
\check{\Sigma}_{f_2} &= \Delta_2 \hat{s}_1 \hat{\sigma}_1 \cdot  (-i\omega_n) J_n, \\
\label{sigmaf3}
\check{\Sigma}_{f_3}
&= \Delta_3  (i\hat{s}_2)\hat{\sigma}_1 \cdot \eta'_n, \\
\label{sigmaf4}
\check{\Sigma}_{f_4} &= \Delta_4  \hat{s}_3 \hat{\sigma}_1 \cdot (-i\omega_n) J_n , \\
\eta'_n
\label{eta'}
&= n_{\mathrm{imp}} v^2_{\mathrm{imp}} \frac{1}{N} \sum_{\bm{k}} \frac{2}{Z} \hspace{10pc} \nonumber \\
&\hspace{3pc} \times \left[ \omega_n^2+c_{\bm{k}}^2-m_{\bm{k}}^2 +V_x^2+V_y^2-V_z^2 \right] , \\
J_n
&= n_{\mathrm{imp}} v^2_{\mathrm{imp}} \frac{1}{N} \sum_{\bm{k}}
\frac{4 m_{\bm{k}}}{Z} .
\end{align}

Before demonstrating $T_c$ under the potential disorder,
we briefly summarize a relation between the self-energy and the pair potential in 
the four cases.
The results in Eq.~(\ref{sigmaf1}) show
that $\check{\Sigma}_{f_1}$ has the same matrix structure 
with the pair potential as shown in Table.~\ref{pair potentials}.
Namely, $\check{\Sigma}_{f_1}$ renormalizes the pair potential $\Delta_1$ which belongs to 
even-frequency spin-singlet even-momentum-parity even-orbital-parity (ESEE) 
pairing symmetry. We will show that this fact explains the robustness
of $\Delta_1$ in the presence of impurity scatterings.
The same feature can be seen in $\check{\Sigma}_{f_3}$ in Eq.~(\ref{sigmaf3}), which 
implies the robustness of $\Delta_3$. 
On the other hand, $\check{\Sigma}_{f_2}$ and $\check{\Sigma}_{f_4}$ have the different 
matrix structure from their pair potentials shown in Table.~\ref{pair potentials}. 
In other words, the impurity self-energy leaves the pair potentials as they are. 
The previous studies suggested that the superconductivity in such cases can be fragile. 
We also note that $\check{\Sigma}_{f_2}$ and $\check{\Sigma}_{f_4}$ enhance the pair 
correlation belonging to
odd-frequency spin-triplet even-momentum-parity even-orbital-parity (OTEE) symmetry. 
In what follows, we discuss characteristic behavior of $T_c$ as a 
function of impurity concentration case by case.

\begin{figure}[tbh]
\begin{center}
\includegraphics[clip,scale=0.7]{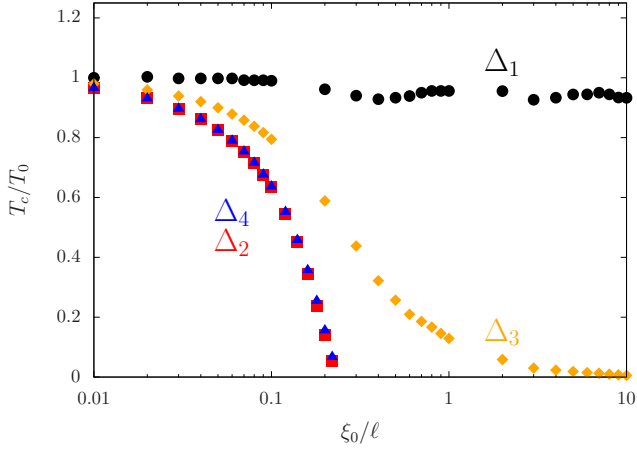}
\end{center}
\caption{
The superconducting transition temperature $T_c$ is plotted as a function of $\xi_0 / \ell$. 
The vertical axis is normalized to the transition temperature in the clean limit $T_0$. 
We fix $T_0$ for all pair potentials.
}
\label{fig:tc}
\end{figure}

\subsubsection{$\Delta_1$}
The gap equation for $\Delta_1$ results in
\begin{align}
\Delta_1 
&= g_1 T \sum_{\omega_n}\frac{1}{N} \sum_{\bm{k}} \frac{\tilde{\Delta}_1}{\tilde{Z}} \hspace{7pc} \nonumber \\
\label{gapeq_d_1}
&\hspace{3pc} \times \left[ \tilde{\omega}_n^2 + \tilde{c}_{\bm{k}}^2 + m_{\bm{k}}^2 
                           + V_x^2+V_y^2+V_z^2 \right].
\end{align}
By comparing with the gap equation in the clean limit in Eq.~(\ref{gapeq1}), 
the renormalized values are defined as
\begin{align}
\tilde{\Delta}_1 &= \Delta_1 (1+\eta_n) , \\
\tilde{Z}(\bm{k},i\omega_n) &= Z(\bm{k},i\tilde{\omega}_n)|_{\mu \rightarrow \tilde{\mu} } , \\
\tilde{c}_{\bm{k}} &= c_{\bm{k}}|_{\mu \rightarrow \tilde{\mu}} .
\end{align}
The impurity self-energy renormalizes 
the pair potential and the Matsubara frequency in the same manner as 
$\Delta_1 \rightarrow \tilde{\Delta}_1$ and 
$\omega_n \rightarrow \tilde{\omega}_n$~\cite{AGD_book}.
We solve the gap equation numerically and plot the transition temperature $T_c$ of $\Delta_1$ 
as a function of $\xi_0 / \ell$ in Fig.~\ref{fig:tc}. 
Here 
$T_0$ is the transition temperature in the clean limit, 
$\xi_0=v_F/(2\pi T_0)$ is the superconducting coherence length, 
$v_F$ is the Fermi velocity, $\ell=v_F \tau_{\mathrm{imp}}$ is the mean free path due to impurity scatterings, 
and $\tau_{\mathrm{imp}}$ is the lifetime of a quasiparticle.
We found that the normal part of self-energy $\check{\Sigma}_g$
is nearly independent of the Matsubara frequency in the low energy region 
for $\omega_n \leq \omega_c$.
Here $\omega_c = 10^3 T_0$ is the cut-off energy of the Matsubara frequency. 
Therefore, we estimate $\tau_{\mathrm{imp}}$ from the imaginary part of $\check{\Sigma}_g$ as
\begin{align}
\frac{1}{\tau_{\mathrm{imp}}} 
= -2 \mathrm{Tr} \left[ \frac{1}{4} \operatorname{Im} \check{\Sigma}_g \right]
\sim 2 \pi \times n_{\mathrm{imp}}v_{\mathrm{imp}}^2 \times 10^{-2} \, [\mathrm{eV}] .
\end{align}
The horizontal axis $\xi_0/\ell$ in Fig.~\ref{fig:tc} is proportional to the 
impurity concentration $n_\mathrm{imp}$. 
The results in Fig.~\ref{fig:tc} show that 
$T_c$ of $\Delta_1$ is almost independent of the impurity concentration as shown with
filled circles.
Such behavior agrees well with $T_c$ in a limiting case of idealistic models.
The previous papers~\cite{Golubov_prb1997,Efremov_prb2011,Asano_prb2018} 
considered two-band superconductivity with the intraband pairing order parameters 
(say $D_1$ and $D_2$) on idealistic two-band electronic structures and
demonstrated that $T_c$ is independent of impurity concentration at $D_1=D_2$.
The interband impurity scatterings disappear in such a symmetric situation, which 
explains the unchanged $T_c$. 
The superconducting state in Cu-doped Bi$_2$Se$_3$ with $\Delta_1$ corresponds to 
the symmetric intraband pairing state in the previous studies. 
In this paper, we 
confirmed that the conclusions of the previous papers on idealistic band structures
are valid even if we calculate $T_c$ on a realistic electronic structure.  
In Fig.~\ref{fig:tc}, the results for $\Delta_1$ show the oscillating behavior.
Although it is not easy to specify the reasons of the oscillations, 
such behavior comes from a realistic band structure. 
In the Born approximation, we conclude that $T_c$ of $\Delta_1$ 
is insensitive to the impurity scatterings.

\subsubsection{$\Delta_3$}
The gap equation for $\Delta_3$ becomes
\begin{align}
\Delta_3 &= 
g_3 T \sum_{\omega_n}\frac{1}{N} \sum_{\bm{k}} \frac{\Delta_3'}{\tilde{Z}} \hspace{7pc} \nonumber \\
\label{gapeq_d_3}
&\hspace{3pc} \times \left[ \tilde{\omega}_n^2 + \tilde{c}_{\bm{k}}^2 - m_{\bm{k}}^2 
                           +V_x^2+V_y^2-V_z^2 \right] , \\
\Delta_3' &= \Delta_3 (1+\eta'_n) .
\end{align}
The pair potential is renormalized by the impurity self-energy as 
$\Delta_3 \rightarrow \Delta_3'$ in Eq.~\eqref{gapeq_d_3} 
in a slightly different way from the relation 
$\omega_n \rightarrow \tilde{\omega}_n$.
By solving Eq.~\eqref{gapeq_d_3}, we plot $T_c$ of $\Delta_3$ as a function of $\xi_0/\ell$ 
in Fig.~\ref{fig:tc}.
The results show that $T_c$ of the spin-singlet interorbital pairing order 
is suppressed slowly with the increase of $\xi_0 / \ell$ 
and goes to zero in the dirty limit. 
A previous paper~\cite{Asano_njphys2018}, however, 
demonstrated on an idealistic two-band structure 
that $T_c$ of a spin-singlet $s$-wave interband pairing 
order is independent of the impurity concentration.
Thus $\Delta_3$ in a Cu-doped Bi$_2$Se$_3$ is more fragile than that in an idealistic 
two-band model.
The difference between the results in the two models can be explained by 
the enhancement of odd-frequency pairing components due to the realistic electronic structures.
The odd-frequency pairing correlation is absent in an idealistic band structure~\cite{Asano_njphys2018}. 
As a result, the impurity self-energy 
renormalizes the pair potential and the Matsubara 
frequency in the same manner, which leads to unchanged $T_c$ versus $\xi_0/\ell$.
In Cu-doped Bi$_2$Se$_3$, on the other hand, 
the asymmetry between two-orbitals ($m_{\bm{k}}$) and the orbital hybridization ($V_z$) 
generate the odd-frequency pairing correlations as described in Eq.~\eqref{f3}.
These correlations contribute negatively to the numerator of the renormalization 
factor of the pair potential $1+\eta'_n$
as shown in $-m_{\bm{k}}^2$ and $-V_z^2$ in Eq.~\eqref{eta'}.
As a consequence, the reduction of the pair potential by odd-frequency pairs 
causes the suppression of $T_c$ in the dirty regime.
We conclude that 
the robustness of the spin-singlet $s$-wave interorbital pairing order depends on 
band structures.

\subsubsection{$\Delta_2$ and $\Delta_4$} 
\label{d2}
The gap equations for $\Delta_2$ and $\Delta_4$ result in
\begin{align}
\Delta_2 
&= g_2 T\sum_{\omega_n} \frac{1}{N} \sum_{\bm{k}} \frac{\Delta_2}{\tilde{Z}} \hspace{7pc} \nonumber \\
&\hspace{1pc} \times \left[ \tilde{\omega}_n^2 + \tilde{c}_{\bm{k}}^2 -m_{\bm{k}}^2 
                           + V_x^2+V_y^2+V_z^2 \right. \nonumber \\
\label{gapeq_d_2}
&\hspace{2pc} \left. - 2 J_n \omega_n \tilde{\omega}_n m_{\bm{k}} \right] , \\
\Delta_4 
&= g_4 T\sum_{\omega_n} \frac{1}{N} \sum_{\bm{k}} \frac{\Delta_4}{\tilde{Z}} \hspace{7pc} \nonumber \\
&\hspace{1pc} \times \left[ \tilde{\omega}_n^2 + \tilde{c}_{\bm{k}}^2 -m_{\bm{k}}^2 
                           + V_x^2-V_y^2+V_z^2 \right. \nonumber \\
\label{gapeq_d_4}
&\hspace{2pc} \left. - 2 J_n \omega_n \tilde{\omega}_n m_{\bm{k}} \right] .
\end{align}
Both $\Delta_2$ and $\Delta_4$ represent spin-triplet interorbital pairing order antisymmetric 
under the permutation of two orbitals.
The numerical results  in Fig.~\ref{fig:tc} indicate that $T_c$ of $\Delta_2$ and that of $\Delta_4$
decrease rapidly with the increase of $\xi_0/\ell$ and vanish around $\xi_0/\ell \approx 0.3$.
The impurity self-energy renormalizes the Matsubara frequency as 
$\omega_n \rightarrow \tilde{\omega}_n$. However, it leaves the pair potentials 
unchanged as shown in Eqs.~\eqref{gapeq_d_2} and \eqref{gapeq_d_4}.
Therefore, $\Delta_2$ and $\Delta_4$ are fragile in the presence of impurities.
The obtained results of $T_c$ for a Cu-doped Bi$_2$Se$_3$ agree even quantitatively with those 
calculated in an idealistic band structure~\cite{Asano_njphys2018}.
The interorbital impurity scatterings mix the electronic states in the two orbitals 
and average the pair potentials over the two orbital degree of freedom. 
As a result, the impurity scatterings wash out the sign of the pair potential 
in Eq.~\eqref{hbdg}, which leads to the suppression of odd-orbital symmetric superconductivity. 
We confirmed that this physical interpretation is valid independent of band structures.

Finally, we compare our results in the present paper with
those in a recent study~\cite{Cavanagh_prb2020}.
The authors of Ref.~\onlinecite{Cavanagh_prb2020} formulated the random impurity 
scatterings based on the two-band picture 
in momentum space, which is obtained by diagonalizing the normal state Hamiltonian 
in the absence of impurities~\cite{Michaeli_prl2012}. 
They mapped a Hamiltonian for an interorbital $s$-wave superconductor with random impurities 
to a Hamiltonian for a single-band unconventional superconductor with random impurities.
As a result, they concluded that $\Delta_2$, $\Delta_3$, and $\Delta_4$ are 
fragile under the potential disorder. 
Their conclusion for $\Delta_3$ does not agree 
with ours obtained by applying the standard method~\cite{AGD_book}. 
The difference in the theoretical methods causes the discrepancy.
A key point might be the self-energy due to interorbital impurity scatterings. 
Actually all of the previous papers~\cite{Golubov_prb1997,Efremov_prb2011,Asano_prb2018,Asano_njphys2018} 
have suggested an importance of the interorbital/interband impurity scatterings on $T_c$. 
Ref.~\onlinecite{Cavanagh_prb2020}, on the other hand, does not consider the interorbital impurity scatterings.

\section{Conclusion}
\label{section:conclusion}

We studied the effects of random nonmagnetic impurities
on the superconducting transition temperature $T_c$ in Cu-doped Bi$_2$Se$_3$.
We consider four types of momentum-independent pair potentials, 
which include the intraorbital pairing ($\Delta_1$), the interorbital-even-parity pairing ($\Delta_3$), 
and the interorbital-odd-parity pairings ($\Delta_2$ and $\Delta_4$).
The effects of the impurity scatterings are considered
through the self-energy of the Green's function within the Born approximation.
$T_c$ of $\Delta_1$ is insensitive to the impurity concentration, 
which is consistent with the previous theories. We find that $\Delta_1$ with 
the electronic structure of a Cu-doped Bi$_2$Se$_3$ corresponds to a limiting case of 
idealistic models~\cite{Golubov_prb1997,Efremov_prb2011,Asano_prb2018}.
$T_c$ of $\Delta_3$ decreases moderately with the increase of impurity concentration 
and vanishes in the dirty limit, which does not agree well with the results 
on an idealistic model~\cite{Asano_njphys2018}. 
The presence of the odd-frequency pairing correlations explain the discrepancy.
$T_c$ of $\Delta_2$ and $\Delta_4$ decrease rapidly 
with the increase of the impurity concentration.
Superconductivity vanishes at a critical value of the impurity concentration.
The results are consistent with those in an idealistic model even quantitatively~\cite{Asano_njphys2018}.

We found that the robustness of the even-orbital-parity order parameters depends on 
the details of the band structures and that 
the odd-orbital-parity order parameters are fragile irrespective of the band structures.

\begin{acknowledgments}
The authors are grateful to P. M. R. Brydon, D. C. Cavanagh, and K. Yada for useful discussions.
This work was supported by KAKENHI (No.~20H01857), 
JSPS Core-to-Core Program (A. Advanced Research Networks), and
JSPS and Russian Foundation for Basic Research under Japan-Russia Research Cooperative Program
Grant No. 19-52-50026.
\end{acknowledgments}

\appendix
\begin{widetext}
\section{Restriction of hopping matrix in tight-binding Hamiltonian}
\label{section:tb hamiltonian}
The crystal structure of Bi$_2$Se$_3$ preserves discrete symmetries~\cite{Zhang_natphys2009,Liu_prb2010} such as 
threefold rotation $R_3$ along the $z$ direction, 
twofold rotation $R_2$ along the $x$ direction, and 
inversion $P$. 
In addition, both the normal and superconducting states 
preserve time-reversal $T$ symmetry.
With the basis of ($\ket{P1_z^+,\uparrow}$, $\ket{P2_z^-,\uparrow}$, $\ket{P1_z^+,\downarrow}$, $\ket{P2_z^-,\downarrow}$),
these symmetry operations can be represented as 
$R_3=\exp(i\frac{\pi}{3} s_3 \sigma_0)$, $R_2=i s_1 \sigma_3$, $P=s_0 \sigma_3$, and 
$T=is_2\sigma_0 \mathcal{K}$, respectively.
Here $\mathcal{K}$ represents the complex conjugation.

Under threefold rotation symmetry, the relation
\begin{align}
\label{t_r3}
\bra{\bm{R},\sigma,s} H \ket{\bm{R}+\bm{a}_i,\sigma',s'}
= \exp \left( i\frac{\pi}{3}(s'_3-s_3) \right) \bra{\bm{R},\sigma,s} H \ket{\bm{R}+\bm{a}_{j},\sigma',s'}, 
\end{align}
is satisfied for $(\bm{a}_i, \bm{a}_j)=(\bm{a}_1, -\bm{a}_2)$, 
$(\bm{a}_2, -\bm{a}_3)$, and $(\bm{a}_3, \bm{a}_1)$.
Under twofold rotation symmetry, the relation
\begin{align}
\label{t_r2}
\bra{\bm{R},\sigma,s} H \ket{\bm{R}+\bm{a}_i,\sigma',s'}
= \sigma'_3 \sigma_3 \bra{\bm{R},\sigma,\overline{s}} H \ket{\bm{R}+\bm{a}_{j},\sigma',\overline{s'}}, 
\end{align}
holds true for $(\bm{a}_{i}, \bm{a}_{j})$ = $(\bm{a}_1, -\bm{a}_3)$, $(\bm{a}_2, -\bm{a}_2)$, 
and $(\bm{a}_3, -\bm{a}_1)$.
As a results of inversion symmetry, we find the relation of
\begin{align}
\label{t_p}
\bra{\bm{R},\sigma,s} H \ket{\bm{R}+\bm{a}_i,\sigma',s'}
= \sigma'_3 \sigma_3 \bra{\bm{R},\sigma,s} H \ket{\bm{R}-\bm{a}_i,\sigma',s'}.
\end{align}
Finally, time-reversal symmetry is described as 
\begin{align}
\label{t_t}
\bra{\bm{R},\sigma,s} H \ket{\bm{R}+\bm{a}_i,\sigma',s'}
= s'_3 s_3 \bra{\bm{R},\sigma',\overline{s'}} H \ket{\bm{R}-\bm{a}_i,\sigma,\overline{s}}.
\end{align}
We have used the notation of
\begin{align}
\sigma_3 &= 
\begin{cases}
+1 & (\sigma=P1_z^+) \\
-1 & (\sigma=P2_z^-)
\end{cases} , \quad
s_3 =
\begin{cases}
+1 & (s = \uparrow) \\
-1 & (s = \downarrow)
\end{cases} , \\
\overline{\sigma} &=
\begin{cases}
P2_z^- & (\sigma=P1_z^+) \\
P1_z^+ & (\sigma=P2_z^-)
\end{cases} , \quad
\overline{s} =
\begin{cases}
\downarrow & (s=\uparrow) \\
\uparrow & (s=\downarrow)
\end{cases} .
\end{align}
According to the conditions in Eqs.~\eqref{t_r3}, \eqref{t_r2}, \eqref{t_p}, and \eqref{t_t}, 
the hopping matrices can be reduced as~\cite{Hashimoto_jpsj2013,Mao_prb2011} 
\begin{align}
\check{t}_{\bm{a}_1} &= \left(
\begin{array}{cccc}
t_{11} & t_{12} & 0 & t_{14} \\
-t_{12} & t_{22} & t_{14} & 0 \\
0 & -t_{14}^{\ast} & t_{11} & t_{12} \\
-t_{14}^{\ast} & 0 & -t_{12} & t_{22}
\end{array}
\right) , \quad
\check{t}_{-\bm{a}_1} = \left(
\begin{array}{cccc}
t_{11} & -t_{12} & 0 & -t_{14} \\
t_{12} & t_{22} & -t_{14} & 0 \\
0 & t_{14}^{\ast} & t_{11} & -t_{12} \\
t_{14}^{\ast} & 0 & t_{12} & t_{22}
\end{array}
\right) , \\
\check{t}_{\bm{a}_2} &= \left(
\begin{array}{cccc}
t_{11} & -t_{12} & 0 & -e^{i2\pi/3}t_{14} \\
t_{12} & t_{22} & -e^{i2\pi/3}t_{14} & 0 \\
0 & -e^{i2\pi/3}t_{14} & t_{11} & -t_{12} \\
-e^{i2\pi/3}t_{14} & 0 & t_{12} & t_{22}
\end{array}
\right) , \quad
\check{t}_{-\bm{a}_2} = \left(
\begin{array}{cccc}
t_{11} & t_{12} & 0 & e^{i2\pi/3}t_{14} \\
-t_{12} & t_{22} & e^{i2\pi/3}t_{14} & 0 \\
0 & e^{i2\pi/3}t_{14} & t_{11} & t_{12} \\
e^{i2\pi/3}t_{14} & 0 & -t_{12} & t_{22}
\end{array}
\right) , \\
\check{t}_{\bm{a}_3} &= \left(
\begin{array}{cccc}
t_{11} & t_{12} & 0 & -t_{14}^{\ast} \\
-t_{12} & t_{22} & -t_{14}^{\ast} & 0 \\
0 & t_{14} & t_{11} & t_{12} \\
t_{14} & 0 & -t_{12} & t_{22}
\end{array}
\right) , \quad
\check{t}_{-\bm{a}_3} = \left(
\begin{array}{cccc}
t_{11} & -t_{12} & 0 & t_{14}^{\ast} \\
t_{12} & t_{22} & t_{14}^{\ast} & 0 \\
0 & -t_{14} & t_{11} & -t_{12} \\
-t_{14} & 0 & t_{12} & t_{22}
\end{array}
\right) , \\
\check{t}_{\bm{a}_4} &= \left(
\begin{array}{cccc}
t'_{11} & t'_{12} & 0 & 0 \\
-t'_{12} & t'_{22} & 0 & 0 \\
0 & 0 & t'_{11} & t'_{12} \\
0 & 0 & -t'_{12} & t'_{22}
\end{array}
\right) , \quad
\check{t}_{-\bm{a}_4} = \left(
\begin{array}{cccc}
t'_{11} & -t'_{12} & 0 & 0 \\
t'_{12} & t'_{22} & 0 & 0 \\
0 & 0 & t'_{11} & -t'_{12} \\
0 & 0 & t'_{12} & t'_{22}
\end{array}
\right) .
\end{align}
In momentum space, the tight-binding Hamiltonian becomes
\begin{align}
\check{H}_N(\bm{k}) &= \left(
\begin{array}{cccc}
c_{\bm{k}} + m_{\bm{k}} & -i( v_3 \alpha_3 (\bm{k}) + v_z \alpha_z(\bm{k}))  & 0 & v(\alpha_y(\bm{k}) + i\alpha_x(\bm{k})) \\
i(v_3 \alpha_3 (\bm{k}) + v_z \alpha_z(\bm{k}) ) & c_{\bm{k}} - m_{\bm{k}} & v(\alpha_y(\bm{k}) + i\alpha_x(\bm{k})) & 0 \\
0 & v(\alpha_y(\bm{k}) - i\alpha_x(\bm{k})) & c_{\bm{k}} + m_{\bm{k}} & -i(v_3 \alpha_3 (\bm{k}) + v_z \alpha_z(\bm{k}) ) \\
v(\alpha_y(\bm{k}) - i\alpha_x(\bm{k})) & 0 & i(v_3 \alpha_3 (\bm{k}) + v_z \alpha_z(\bm{k})) & c_{\bm{k}} - m_{\bm{k}}
\end{array}
\right),
\end{align}
with
\begin{align}
c_{\bm{k}} &= -\mu + c_1 \alpha_1 (\bm{k}) + c_2 \alpha_2 (\bm{k}) , \quad
m_{\bm{k}} = m_0 + m_1 \alpha_1 (\bm{k}) + m_2 \alpha_2 (\bm{k}) , \\
c_1 &= -\frac{c^2}{2} (t'_{11} + t'_{22}) , \quad
c_2 = -\frac{3a^2}{4} (t_{11} + t_{22}) , \quad
\mu = -3(t_{11}+t_{22}) - (t'_{11}+t'_{22}) -\varepsilon , \\
m_1 &= -\frac{c^2}{2} (t'_{11}-t'_{22}) , \quad
m_2 = -\frac{3a^2}{4} (t_{11}-t_{22}) , \quad
m_0 = 3(t_{11}-t_{22}) + t'_{11} - t'_{22} , \\
v &= -3ie^{i2\pi/3} a t_{14} , \quad
v_z = -2 c t'_{12} , \quad
v_3 = \frac{3a^3}{4} t_{12}.
\end{align}
Here
$\alpha_1 (\bm{k})$, $\alpha_2 (\bm{k})$, $\alpha_x (\bm{k})$, $\alpha_y (\bm{k})$, and $\alpha_z (\bm{k})$, 
are defined in Eqs.~\eqref{a1}-\eqref{az}. 
We also define $\alpha_3 (\bm{k}) = -\frac{8}{3a^3} (2\cos \frac{\sqrt{3}}{2} k_x a \sin \frac{1}{2} k_y a -\sin k_y a)$. 
In this paper, we set the parameters as follows~\cite{Hashimoto_jpsj2013,Mizushima_prb2014}: 
$a=4.14~\mathrm{\AA}$, $c=28.7~\mathrm{\AA}$, 
$\mu=0.5~\mathrm{eV}$, $c_2=30.4~\mathrm{eV\AA^2}$, $m_0=-0.28~\mathrm{eV}$, $m_2=44.5~\mathrm{eV\AA^2}$, 
$v=3.33~\mathrm{eV\AA}$, $c_1/c^2=0.024~\mathrm{eV}$, $m_1/c^2=0.20~\mathrm{eV}$, and $v_z/c=0.32~\mathrm{eV}$. 
We choose $v_3 = 0$ for simplicity~\cite{Zhang_natphys2009,Liu_prb2010}.

\section{Unitary equivalence of the Hamiltonian with intraorbital pairing order}
\label{section:eq between 1a & 1b}
The superconducting state with 
$s$-wave spin-singlet intraorbital pairing order is described 
by a following Bogoliubov-de Gennes Hamiltonian~\cite{Asano_prb2018}.
\begin{gather}
\label{hintra}
\bar{H}^{(0)}_{\bm{k}} ( \theta , \varphi_1 , \varphi_2 ) =
\left[\begin{array}{cccccccc}
\xi_1 & -iV_z e^{i\theta} & 0 & V e^{i\theta} & 0 & 0 & \Delta_{P1_{z}^+} & 0 \\
iV_z e^{-i\theta} & \xi_2 & V e^{-i\theta} & 0 & 0 & 0 & 0 & \Delta_{P2_{z}^-} \\
0 & V^{\ast} e^{i\theta} & \xi_1 & -iV_z e^{i\theta} & -\Delta_{P1_{z}^+} & 0 & 0 & 0 \\
V^{\ast} e^{-i\theta} & 0 & iV_z e^{-i\theta} & \xi_2 & 0 & -\Delta_{P2_{z}^-} & 0 & 0 \\
0 & 0 & -\Delta_{P1_{z}^+}^{\ast} & 0 & -\xi_1 & iV_z e^{-i\theta} & 0 & V^{\ast} e^{-i\theta} \\
0 & 0 & 0 & -\Delta_{P2_{z}^-}^{\ast} & -iV_z e^{i\theta} & -\xi_2 & V^{\ast} e^{i\theta} & 0 \\
\Delta_{P1_{z}^+}^{\ast} & 0 & 0 & 0 & 0 & V e^{-i\theta} & -\xi_1 & iV_z e^{-i\theta} \\
0 & \Delta_{P2_{z}^-}^{\ast} & 0 & 0 & V e^{i\theta} & 0 & -iV_z e^{i\theta} & -\xi_2
\end{array} \right] , \\
\xi_1 = c_{\bm{k}}+m_{\bm{k}} ,\quad
\xi_2 = c_{\bm{k}}-m_{\bm{k}} ,\quad
V = v(\alpha_y (\bm{k})+i\alpha_x (\bm{k})) ,\quad
V_z = v_z \alpha_z (\bm{k}) , \\
\Delta_{P1_{z}^+} = \frac{g_{P1_{z}^+}}{N} \sum_{\bm{k}}
\langle \psi_{P1_{z}^+,\uparrow} (\bm{k}) \psi_{P1_{z}^+,\downarrow} (-\bm{k}) \rangle
=|\Delta_{P1_{z}^+}| e^{i\varphi_1} , \\
\Delta_{P2_{z}^-} = \frac{g_{P2_{z}^-}}{N} \sum_{\bm{k}}
\langle \psi_{P2_{z}^-,\uparrow} (\bm{k}) \psi_{P2_{z}^-,\downarrow} (-\bm{k}) \rangle
=|\Delta_{P2_{z}^-}| e^{i\varphi_2} ,
\end{gather}
where $g_{\sigma}>0$ represents the attractive interaction between two electrons in the orbital $\sigma$ 
and $\theta$ denotes the phase of the hybridization in the normal state.
We obtain the normal part of $\bar{H}^{(0)}_{\bm{k}} ( \theta,\varphi_1,\varphi_2)$ from Eq.~\eqref{hn} 
by choosing
$\psi_{P1_{z}^+,s} \, \rightarrow \, \psi_{P1_{z}^+,s} e^{i\theta/2}$ and 
$\psi_{P2_{z}^-,s} \, \rightarrow \, \psi_{P2_{z}^-,s} e^{-i\theta/2}$.
Although the phase factor $e^{i\theta}$ does not affect the physics in the normal state,
 such a gauge transformation affects 
the relative phase difference between the order parameters $\varphi_1-\varphi_2$~\cite{Asano_prb2018}.

Time-reversal symmetry of $\bar{H}^{(0)}_{\bm{k}}$ is represented by
\begin{align}
\mathcal{T} \bar{H}^{(0)}_{\bm{k}} \mathcal{T}^{-1} = \bar{H}^{(0)}_{-\bm{k}} , \quad
\mathcal{T} = \hat{\tau}_0 (i\hat{s}_2) \hat{\sigma}_0 \mathcal{K} .
\end{align}
If we find a transformation $\bar{U}$ which eliminates all the phase factors in Eq.~\eqref{hintra},
it is possible to show time-reversal symmetry of $\bar{H}^{(0)}_{\bm{k}}$~\cite{Asano_prb2018}.
By applying the unitary transformation, 
\begin{align}
\bar{U}={\mathrm{diag}}[e^{-i\varphi_1/2} , e^{-i\varphi_2/2} , e^{-i\varphi_1/2} , e^{-i\varphi_2/2} ,
e^{i\varphi_1/2} , e^{i\varphi_2/2} , e^{i\varphi_1/2} , e^{i\varphi_2/2} ] ,
\end{align}
the Hamiltonian is transformed into
\begin{align}
\bar{U} \bar{H}^{(0)}_{\bm{k}} (\theta,\varphi_1,\varphi_2) \bar{U}^{\dag} = 
\bar{H}^{(0)}_{\bm{k}} (\theta-\frac{\varphi_1-\varphi_2}{2} , 0 , 0) .
\end{align}
Therefore, the three phases must satisfy a relation
\begin{align}
2\theta - \varphi_1 + \varphi_2 = 2 \pi n ,
\end{align}
with $n$ being an integer
for the Hamiltonian to preserve time-reversal symmetry.
By tuning $\theta=0$ at $n=0$, the two pair potentials have the same sign 
with each other because of $\varphi_1-\varphi_2=0$. 
By tuning $\theta=\pi/2$, on the other hand, $\bar{H}^{(0)}_{\bm{k}} (\pi/2,0,\pi)$ describes a state 
where two pair potentials have the opposite sign to each other. 
It is easy to show that $\bar{H}^{(0)}_{\bm{k}} (\pi/2,0,\pi)$ and 
$\bar{H}^{(0)}_{\bm{k}} (0,0,0)$ are unitary equivalent to each other.
We set $g_{P1_{z}^+}=g_{P2_{z}^-}=g_1$ and 
$\Delta_{P1_{z}^+} = \Delta_{P2_{z}^-} = \Delta_1$ in section~\ref{sectoion:model}.
Under the condition, $\Delta_1 (i\hat{s}_2)\hat{\sigma}_3$ is 
unitary equivalent to $\check{\Delta}_1=\Delta_1 (i\hat{s}_2)\hat{\sigma}_0$.
\end{widetext}

\bibliography{paper_submit}

\end{document}